# Blind Multi-user Detection for Autonomous Grant-free High-Overloading MA without Reference Signal

Zhifeng Yuan*, Yuzhou Hu, Weimin Li, Jianqiang Dai
* The corresponding author, email: yuan.zhifeng@zte.com.cn
ZTE Corporation, South Keji Road, 55, Shenzhen, China, 508118

## ABSTRACT

**In this paper, a novel blind multi-user detection(MUD) framework for autonomous grant-free(AGF) high-overloading non-orthogonal multiple access(NOMA) is introduced in detail aimed at fulfilling the requirements of 5G massive Machine Type Communications(mMTC). From the perspective of the transmitter side, pros and cons regarding diverse types of emerging grant-free transmission, particularly autonomous grant-free(AGF), are elaborated and presented in a comparative manner. In the receiver end, codeword-level successive interference cancellation (CL-SIC) is revealed as the main framework to perform MUD. In addition, underpinning state-of-art blind ideas such as blind activation detection taking advantage of the statistical metric of the aggregate signals, blind equalization based on the constellation's simple geometric character of low order modulation symbols, and blind channel estimation employing solely the successfully decoded codewords are explained.**

.

***Keywords: massive Machine Type Communications (mMTC); 5G; non-orthogonal multiple access(NOMA); grant-free;one-shot;blind multiple user detection(MUD); blind activation detections; blind equalization; data pilot;data-assisted channel estimation***

## I. INTRODUCTION

Machine-type-communication(MTC) is widely anticipated to be a very important scenario in the future generations of wireless network[1-3]. MTC can be divided into two main types, i.e., massive MTC(mMTC) with low data rate, and MTC with low latency and high reliability. In mMTC scenario, it has been forecasted that a massive number of devices(UEs) transmitting sporadic small data packets will connect to the networks. One type of KPI about connection density of mMTC is $10^6/km^2$[4]. Hence, the following two capabilities are very crucial for an affordable mMTC network deployment, i.e., consisting of low-cost, power-saving devices and supporting massive infrequent small packets with reasonable spectrum. As mMTC is a new scenario, the existing wireless networks naturally were not designed for it and we would see they are indeed ill suited for the mMTC traffic.

To enforce grant-based data transmission, which is a fundamental feature of existing wireless networks such as LTE, a device needs to send the scheduling request and wait for dynamic grant prior to each data packet transmission[5]. Moreover, in order to support massive connections with limited system resources and keep the devices' power consumption as low as possible, each device had better release its connection session and turn to deep-sleep/idle mode once its data transmission is finished. To access the network again, these idle devices have to perform additional and usually more complex contention-based random access firstly. Intuitively, all these closed-loop procedures before a small data packet transmission could consume even more resources and energy than the data itself, leading to very low efficiency of spectrum and energy per mMTC device, thus block an economic deployment of mMTC network. New multiple access(MA) strategies and techniques which can accommodate the traffic characteristics of mMTC are consequently called for.

Grant-free data transmission, in contrast, means that devices can directly transmit data without the need to send scheduling request and wait for dynamic grant, leading to a much simplified transmitting procedure compared to the grant-based approach. It's important to clarify that grant-free can be either pre-configured or autonomous and these two types of grant-free have distinctly different merits and demerits, making them suitable to different scenarios. In the former, each UE is pre-configured by the BS statically or semi-statically UE-specific regular or periodic physical resources[5]. Then, each UE can transmit data directly on its resources without dynamic grant. Thanks to the pre-configuration performed by the base station(BS), the UE-specific physical resources and MA signatures such as spreading sequence, preamble and demodulation reference signal(DMRS) can be easily made either orthogonal or low correlated. Orthogonal pre-configuration can naturally much simplify the receiver and ensure a more robust performance due to the interference-free links, nevertheless at the expense of lower spectral efficiency or connection density. On the other hand, low correlated pre-configuration enables the system to accommodate more connections than the orthogonal method, with the penalty of performance degradation per link or a more complex MUD receiver in the BS, due to the inter-user interference. Strictly speaking, this type of grant-free is only dynamic-grant-free rather than fully grant-free, because it still needs static pre-configuration or static grant. It may work well for regular traffic such as periodic traffic without cell switching[6], because a traffic-matching regular resource pattern[7] can be preconfigured to avoid the waste of physical resources.

However, it is much less efficient for irregular traffic such as event-driven non-periodic traffic[8] or mobile traffic involving cell switching. The former traffic would incur serious waste of resource or very large latency as matching resource allocations for such irregular traffic is hard if not impossible, and the latter requires complicated reconfiguration once cell switching happens[9]. Also, retransmission and the support of flexible packet size in pre-configured grant-free transmission are nontrivial. What's more, things would be worse for the pre-configured grant-free if the number of devices with irregular traffic were very large, which could be the typical situation of mMTC[10-11]. In this case, MA signature assignment and maintenance would be genuinely challenging.

On the other hand, in the autonomous grant-free(AGF) transmission, UE-specific pre-configuration is no longer necessary. Therefore neither dynamic grant nor static/semi-static grant for a particular device is required and fully grant-free is achieved. Each device can transmit data directly in deep-sleep state by randomly selecting physical resources from a cell-specific or system- specific resource pool, and turn to deep-sleep again once the transmission is finished to save its battery consumption. Considering devices' random selection of physical resources and MA signature, the AGF can also be dubbed as random grant-free or contention-based grant-free, in contrast to the alias determined grant-free or contention-free grant-free corresponding to the pre-configuration case.

Free from the constraints of pre-configuration, revolutionary AGF can extremely simplify the transmitter as well as the network compared to the grant-based and the pre-configured grant-free cases in the following aspects,

- ✧ eliminating all the closed-loop procedures before data transmission
- ✧ avoiding the pre-configuration and re-configuration
- ✧ simplifying the retransmission
- ✧ support of various packet size, etc.

This simplification is particularly useful for the system designated to accommodate massive connections each with sporadic small data packet in the sense that the overhead needed to perform several closed-loop procedures before every small packet or the pre-configuration/re-configuration would be too expensive for such system.

From the view of receiving side, receiver for grant-based is relatively simple because all transmissions are controlled by the BS, and the information needed for demodulation can be easily acquired exploiting certain reference signal(RS), etc. resulting into several single-user non-blind detections. When it comes to the receivers for grant-free, including both the pre-configured grant-free for irregular traffic and AGF, blind detection is necessary. Naturally the pre-configuration information would lay off some blind detection burden thanks to the reduction of hypothetical search range and the elusion of crucial MA signature collision problem. However, when it comes to AGF, nothing is known a priori at the BS, thus a fully blind MUD exhausting all hypotheses is intuitively required. What's more, incontrollable inter-user interference and collision would be inevitable with random MA signature selection, making the design of blind MUD receiver more challenging. In the academics, such MA signature as preamble/DMRS collision has been widely regarded as the bottleneck to the performance of MTC and it's also dubbed as user activity detection, preamble detection/estimation, random access overload etc in diverse contexts[12-15]. Efficient handling of the incontrollable inter-user interference and MA signature collision in the design of blind MUD is supposed to be the key for workable high overloading AGF transmission.

Generally speaking, blind MUD for AGF could rely on specific RS such as preamble[16-18], or the data itself solely(a.k.a data-only blind MUD, which is the emphasis of this paper). Preamble-based blind MUD is usually easier because the blind detection burden mainly comes from the preamble detection, involving exhaustive correlations of all possible known preambles. However, collision induced preamble detection related issues still exist especially in high-overloading scenario, where data-only scheme, being inherently collision-proof, exhibits non-negligible advantages. Further analysis shall be carried out in the following section.

In this paper, we discuss the design principles of the transceiver for the more challenging yet valuable AGF for mMTC scenario while proposing a generic blind MUD receiver accordingly. The paper is organized as follows. In Section II, the potentials of high-overloading autonomous grant-free are discussed first and then a transceiver from the MUSA(multi-user shared access)[23] scheme is taken as an example to take advantage of the potentials. Three key techniques of the unique data-only blind MUD, including blind activity detection, blind equalization and data-pilot based channel estimation in the reconstruction, are presented in Section III. In Section IV, performance evaluations are offered. The conclusions are provided in Section V.

## II. Potentials and Transceiver for autonomous grant-free High-Overloading Multiple Access

### A. *Potentials of autonomous grant-free high-overloading*

As discussed above, blind MUD for AGF transmission is more complicated than its counterpart for grant-based or pre-configured grant-free. Further, blind MUD achieving high overloading in AGF is much more challenging as a result of the inter-user interference and the fact that the more critical MA signature collision problem would increase rapidly with the loading as follows,

$$P_{c,c} = 1 - \frac{A_N^M}{N^M} \tag{1}$$

, where $P_{c,c}$ is the collision probability, M represent the number of UEs randomly selecting MA signatures from the pool of size N, $A$ is the full permutation operator

Nevertheless, potential discriminations among multiplexed UEs committing AGF transmissions exist in the following four domains. Compared to preamble-based scheme, data-only MUD would provide at least two more additional discrimination domains, i.e. power domain and constellation domain of low order modulated symbols.

- Power domain
- Code domain
- Spatial domain
- Constellation domain of low order modulated symbols

Preamble-based scheme could ease some blind detection burden. If there is no preamble collision, the blind activity detection can be accomplished by classic correlation peak detection and further channel estimation for these detected active devices can be easily performed after the preamble detection[19], based on which a non-blind MUD could demodulate the information bits of all UEs. However, if preamble collision happens, the preamble-detector based on correlation peak detecting could not discover such collision most of the time even there is power disparity in the collided preamble, leading to miss detections or superposed channel estimation of collided UEs and a subsequent severely degraded performance[20]. The problems of miss detection and inaccurate channel estimation become more severe with the increasing system load, thus clip the system capacity to a great extent. Also, the preamble generally consumes non-negligible overhead in the context of small data packet transmission, leaving a much reduced amount of physical resources for data than the data-only scheme.

When it comes to data-only scheme, if two or more data-only packets of disparate received powers superpose directly in a collision fashion, they can still be separated potentially in the following manner. Decode and strip off the stronger packet from the aggregate received signal first and then decode the remaining weaker packet. It should be noted that received power disparity among the UEs is an inherent feature in AGF transmission of sporadic small packet since no closed-loop power control is performed and the fast fading induced power disparity resides with open-loop power control. More disparity in the received powers of the superposed data packets provides better chance of discriminating them. After the stronger UE 's data packet is decoded correctly, the reconstructed symbols can be used to refine its channel estimation. Because the UEs' data are independent of each other, the channel estimation of the stronger UE would be rather accurate even in case of the data packet collision. Thus the stronger data packet can be subtracted from the superposed signal with little residual error. This is the so-called data-pilot technique[21]. It should be noted that, the procedure of decoding and subtracting successively is the main characteristic of the codeword-level SIC(CL-SIC) receiver. The power disparity could be leveraged to provide certain discrimination for the purpose of collision handling by means of CL-SIC and data-pilot, but it is still not enough for high-overloading system.

It can be inferred from the collision probability formula (1) that larger pool size N is beneficial to collision reduction. Spreading with non-orthogonal sequences can be leveraged to reduce the collision probability in the sense that non-orthogonal sequences of a given length bearing a penalty of some level of inter-code interference would outnumber greatly the sequence length itself which equals the number of equivalent orthogonal partitions of the time-frequency resources. If higher level of inter-code interference is allowed, more spreading sequences could be found of a given length, subject to the theoretical limit of Welch Bound[22]. Better performance could be anticipated as direct collision is replaced by possible inter-UE interferences subject to pre-set bound. In other words, non-orthogonal spreading technique allows a more graceful performance degradation of the AGF system than the collision induced harsh performance 'hard landing' in non-spreading schemes when system loading increases. Therefore a well-designed non-orthogonal spreading code set is critical to higher connection density or higher system loading for AGF. In the context of spreading, system loading is defined as the number of simultaneously accessing UEs to the length of spreading code, and it's conventionally called 'overloading' if the number of simultaneously accessing UEs is larger than the length of spreading code[23].

Third, if multiple receiving antennas are installed at the BS, the spatial domain discrimination, originating from the independent spatial channels, could be exploited to separate the collided data packets. Blind MUD exploiting spatial domain discriminative potential will be discussed in our another paper. Here we confine ourselves to the single antenna case.

Fourth, constellation domain discrimination associated with low-order modulated symbols can be employed to separate the collided data. Take BPSK as an example, the modulated symbols occupy only one dimension of the complex plane. If two collided data packets experience geometrically perpendicular physical channels in the complex plane, then the received BPSK symbols of them are also perpendicular subject to AWGN contamination. Therefore, conventional demodulation of one BPSK packet would not be interfered by the other. Concrete discussions shall be carried out in section III..

In summary, an AGF scheme achieving high overloading would be highly desirable for mMTC. This paper is devoted to the fundamentals of designing such a scheme exploiting the discriminations in the three domains above.

### *B. Data-only Transceiver Design Principal*

In this subsection, we strive to reveal several transceiver design principles aimed at transforming the above potentials into realistic high-overloading system capacity. At transmitter side, first, data-only transmission coupled with non-orthogonal sequence spreading should be exploited. Additionally, to reserve enough margin for random inter-user interference of grant-free and fulfill the coverage enhancement requirements of mMTC application scenario, relatively low MCS is suitable choice. Therefore BPSK or QPSK modulation of the transmit symbols is highly likely to be supported. We will limit our discussion within the scope of BPSK hereafter and the algorithms could be easily extended to QPSK case .It should be noted that a simple open-loop downlink synchronization needs to be performed prior to the AGF transmission and the cyclic

prefix(CP) duration of OFDM symbol should be long enough so that the uplink signals of multiple UEs can be roughly synchronized within the CP at the receiver.

At receiver side, posterior to OFDM demodulation operations, CL-SIC, combined with data-pilot is most desirable in terms of fulfilling potential high-overloading in the context of AGF. Theoretically, MMSE-SIC receiver with channel information is optimal in terms of capacity achieving for multi-user detection: it “implements” the chain rule of mutual information[21]. As to Blind MUD for spreading AGF, MMSE-SIC is also near optimal out of the following reasons,

- ✧ MMSE is information lossless, and MMSE de-spreading not only minimizes the mean square error but also maximizes the SNR of the de-spread signal[21]
- ✧ With successfully decoded stronger packets stripped off from the aggregate signal, the performance of decoding weaker packets would be much improved.
- ✧ The decoding and subtracting can separate the collision much more efficient than the approach utilizing soft information.[26]
- ✧ The decoded codeword can be use to refine channel estimation., leading to reduced residual error.

As a matter of fact, the data-only receiver introduced in the following subsection owes its design inspiration to this type of receiver consisting of MMSE detection module and SIC related modules.

### *C. Data-only Transceiver Architecture*

It is indeed challenging to design a transceiver architecture following the above principles. As a matter of fact, when it comes to decoding the stronger UE first in data-only scheme , no reference signals could be exploited to determine which packet is stronger and to perform subsequent channel equalizations. In the latter part of this subsection, some state-of-art blind ideas would be proposed to address these difficulties.

Prior to that, some further discussions on spreading sequences are necessary for the fulfillment of the transmitter side. Blind SIC for high-overloading is highly correlated with the spreading sequence, as has been discussed in [23-24]. Given loading rate, longer sequence length means more UEs transmitting superposed spread data packets. For example, 300% overloading means 12 superposed UEs for length-4 sequences while 192 UEs for length-64 sequences. Blind SIC for 192 UEs is much harder than that of 12 UEs in the following aspects,

- ✧ The effect of error propagation to the weaker UEs even with data pilot still can hardly be ignored if the number of UEs is large.
- ✧ The delay of SIC is increased as the number of UEs grows.
- ✧ To achieve better performance, MMSE rather than MF de-spreading may be necessary. MMSE de-spreading requires the inversion of covariance matrix whose complexity is proportional to the length of spreading sequence.
- ✧ Blind spreading sequence detection and the MMSE de-spreading function better when the signals in the range of sequence length experiences a coherent channel. But longer sequence occupying more time-frequency resources is more likely to exceed the coherent time or bandwidth, particularly in scenarios where certain time/frequency offsets(To/Fo) exist.

In short, to achieve a certain loading level, shorter spreading sequence means fewer simultaneously accessing UEs, making it a preferred option from the perspective of SIC-type blind MUD .

Nevertheless, the following issues need to be considered to fully realize the potentials of short sequences:

- ✧ The number of low-correlated short sequences is constrained by the Welch Bound. Therefore more efforts should be paid to design a code set accommodating more sequences with lower cross-correlation.
- ✧ Ambiguity could happen in blind MUD: i.e., the sequence used to de-spread the successfully decoded stream may not be that originally selected by the UE. This biases the reconstructed symbols and could consequently degrade the performance of the succeeding UEs. It could be solved by containing the sequence information into the CRC coded bits as depicted in Figure 1.
- ✧ Coverage ability is reduced with short sequence. Low MCS especially low modulation order can be applied to enhance coverage. Further, two-stage spreading with short sequences can be used to reach similar coverage as that achieved by equivalent long sequences.

Taking the above ideas into account, such short sequences as Appendix A are well designed in the following sense. First it's optimal in terms of the mean square of the cross correlations among the spread sequences. Second, the computational complexity is reduced for receiver when 2-tuple complex sequences consisting of elements from the set[1,-1,j,-j] is employed. Thanks to the features of these sequences, complex number multiplication in the despreading and data-pilot module could be transformed into addition of real and complex parts. The spreading based transmitter architecture is illustrated in Figure 1. The index information of the sequence selected by the UE could either occupy several bits' overhead in the information bits or determined by the information bits following a certain pattern. In this way, BS would know the sequence selected by the UE once the data is decoded successfully, which is critical for the reconstruction in the receiver end.

At receiver, blind sequence detection, blind MMSE estimation together with blind equalization component collectively corresponds to the MMSE estimation module while its data-pilot based joint channel estimation and the following 2 components as illustrated in correspond to the SIC related module. Despite the above similarity, it differs from traditional MMSE-SIC receiver at least in the following 3 key components. First is the blind sequence detection component, where a metric is proposed to measure the likelihood that each of the available sequences is selected by the UEs. Further analysis demonstrates the inherent connections between the metric and the post-SINR of the data stream for any given available sequence. Post-SINR is defined as the hard detection SINR calculated from the data stream posterior to blind MMSE detection employing the given available sequence and blind equalization procedure. Second is the blind equalization component, as BS has no knowledge of channel information at all, the component takes full advantage of the geometric feature of the constellation map of BPSK symbols to enforce blind equalization to the blind MMSE detected data stream and strives to obtain a resultant near MMSE estimated data stream with channel and sequence information known at the BS. The blind equalized data stream risks a phase ambiguity of $\pi$, which is why the inverse of it should also be decoded(bi-stream decoding) in case that the original fails the CRC check. Third is the data-pilot based joint channel estimation component, decoded data plays the role of pilot and assures an accurate channel estimation for data stream reconstruction thanks to the low correlation of transmit data streams among multiple users. The reconstructed data is then subtracted from the received signal. In the following, we will elaborate the 3 components one by one together with another key component.

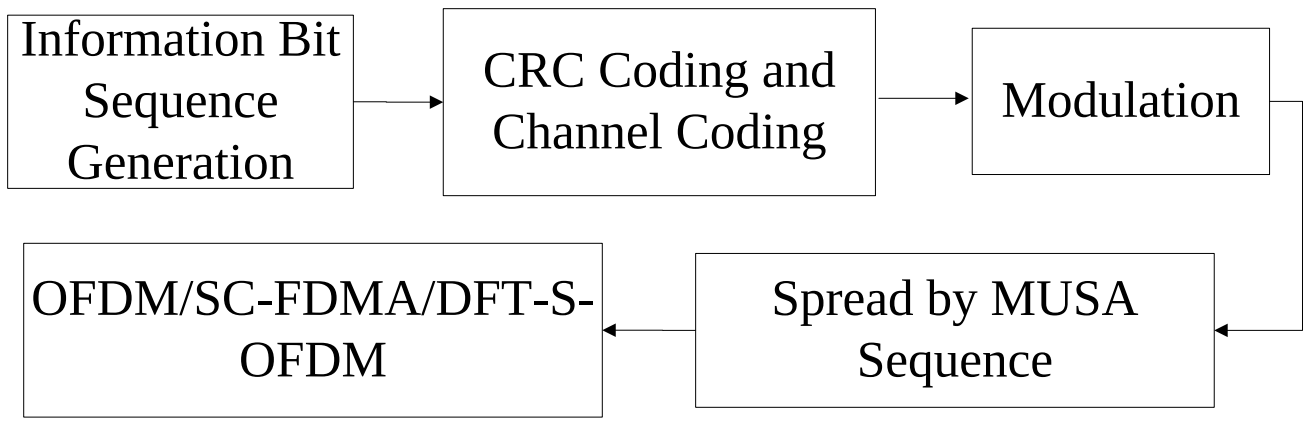

**Figure 1 Data-Only Transmitter**

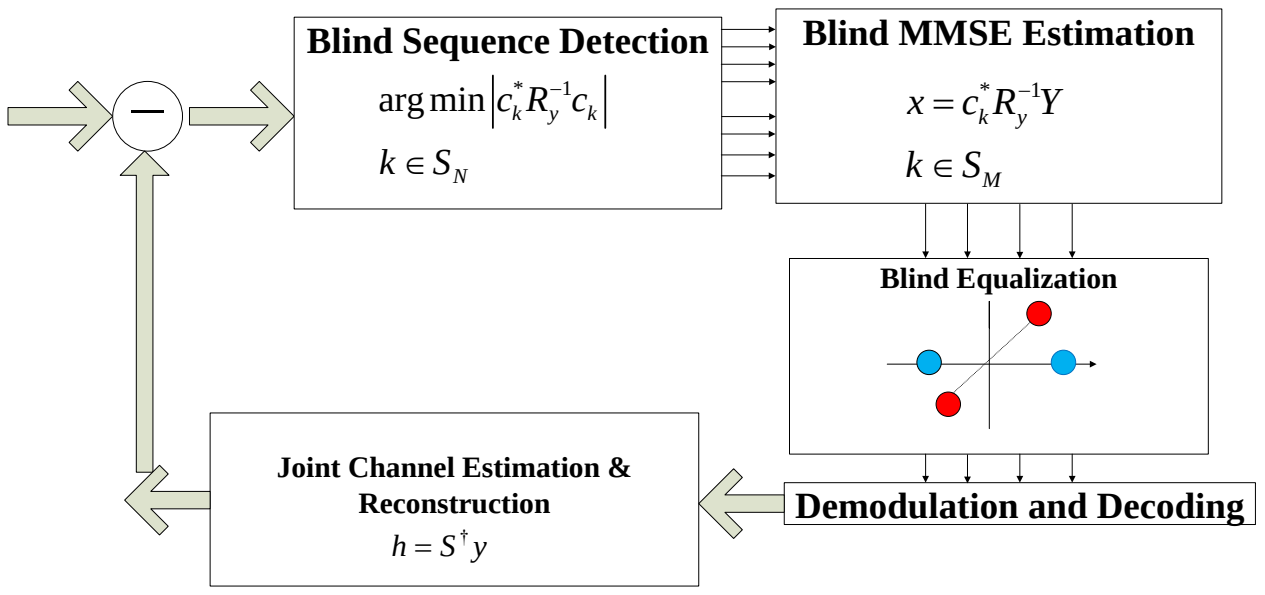

**Figure 2 Data-Only Receiver**

### *D. Data-only Transceiver System Model*

According to the transceiver architecture depicted in the previous subsection, each BPSK from UEs is first spread by their respective spread sequences and then multiplexed on the same resource elements(REs) composing the spread unit before proceeding into the OFDM transmission module. In this paper, we focus on the symbol-level spreading scheme whose spread unit consists of adjacent OFDM symbols on the same subcarrier. The spread unit-wise transceiver equation which views multiplexed UEs other than $u$ as interference can be modeled as follows,

$$y = h_u s_u + n_u \tag{2}$$

, where $s_u$ is the transmit symbol from UE $u$ prior to sequence spreading, $y \in C^{L\times 1}$ is the spread-unit wise received superposed signal from all the UEs, $h_u \in C^{L\times 1}$ is the equivalent channel of UE $u$, and $n_u \in C^{L\times 1}$ denotes the interference plus noise faced by UE $u$. Note that the sequence together with the channel is absorbed into the equivalent channel $h_u$ as follows,

$$h_u = c_u \Theta g_u \tag{3}$$

, where $c_u \in C^{L\times 1}$ is the sequence selected by UE $u$, $g_u \in C^{L\times 1}$ denotes the frequency domain channel and □ is the element-wise product operation. $n_u$ includes the interference from the other M-1 UEs multiplexed on the same REs as UE u and the additive white Gaussian noise at the receiver and can be expressed as follows,

$$n_u = \sum_{i=1, i\neq u}^{M} h_i s_i + n_{awgn} \tag{4}$$

Upon adoption of short spread sequences, the elements of $g_u$ are similar to each other both in amplitude and in phase if we don't consider very large To/Fo. Therefore in the following we may use without mentioning the approximate equation on equivalent channel for simplicity of expression,

$$h_u \doteq c_u g_u \tag{5}$$

, where scalar $g_u$ is the channel of either of the $L$ consecutive symbols. In the performance evaluation section, we consider flat fading channel case where channel stays invariant across the spread PRB, making the approximate equation the following exact equation,

$$h_u = c_u g_u \tag{6}$$

## III. KEY COMPONENTS OF DATA-ONLY BLIND MUD

Apart from the common components in the conventional MMSE-SIC MUD architecture. Data-only blind MUD possesses 4 key components, as illustrated in Figure 2 respectively blind sequence detection, blind MMSE estimation, blind equalization as well as joint channel estimation and reconstruction which shall be elaborated in the following subsections.

### *A. Blind Activity estimation based on the MMSE metric*

Conventional MMSE-SIC receiver requires the knowledge of channel information in its MMSE estimation module. However, data-only receiver has access to neither spread sequence selected by any UE u nor the corresponding physical channel, constituting equivalent channel experienced by the UE as introduced in the transceiver system model. To address this, data-only receiver could divide its MMSE estimation module into 2 major components, i.e. blind sequence detection coupled with blind MMSE estimation and blind equalization. Intuitively speaking, data-only receiver should first determine the sequences most likely to be selected by the UEs based on which a preliminary MMSE estimation could be performed. The resultant data stream should be further equalized where the physical channel experienced by each UE could be compensated and for BPSK modulation, this additional equalization corresponds to the rotation and scaling of 2 clusters of scatter points in the complex plane to the clusters centered on the {+1, -1} transmit symbols.

This subsection is devoted to the discussion of the underpinning theory and methodology of the first component which could be dubbed as either blind activity detection or blind sequence detection. In the following, we would call the component blind sequence detection. The blind sequence detection could be performed in a direct yet complexity expensive manner as follows. First, we could apply each of the spread sequences in the selection pool to the received data stream, procuring a pool of equal amount of un-equalized data streams $\hat{x}_{MMSE_b}$. Then blind equalization techniques to be introduced in the next section could be applied to deprive all the un-equalized data streams $\hat{x}_{MMSE_b}$ of the channel distortion(including To/Fo biased channel). Finally we could decode the several best data streams in terms of hard detection SINR. The complexity is even more expensive in realistic scenario with To/Fo biased channel whose estimation and compensation needs demanding multiplication operations.

For the purpose of computation complexity reduction, a preprocessing procedure could be added evaluating the available spread sequences and ruling out the irrelevant sequences as possible based on certain criteria prior to proceeding into the following blind MMSE estimation and blind equalization procedures. As the data stream with higher post-SINR is more likely to be successfully decoded compared with other data streams. The evaluation metric better reflecting the post-SINR of data stream procured from a given spread sequence will bring out better performance.

In this paper, we offer a data-only blind sequence detection metric as follows,

$$\begin{aligned} M_k &= c_k^* R_y^{-1} c_k \\ R_y &\overset{\Delta}{=} E(yy^*) \end{aligned} \tag{7}$$

,where $c_k$ is the $k_{th}$ sequence in the selection pool, and $R_y$ is the correlation matrix of the aggregate spread signal..in vector form as in formula(2).

Further deductions on the relationship between $M_k$ and the post SINR of the un-equalized de-spread data stream procured from $c_k$, denoted as $\gamma_k$, is presented as follows,

$$\begin{aligned} \gamma_k &= h_k^* \left( \sum_{i=1, i\neq k}^{M} h_i h_i^* + \sigma^2 I \right)^{-1} h_k \\ &= c_k^* \left( \sum_{i=1, i\neq k}^{M} h_i h_i^* + \sigma^2 I \right)^{-1} c_k \left\| h_k \right\|^2 \\ &= x_{-k} \left\| h_k \right\|^2 \end{aligned} \tag{8}$$

, where

$$x_{-k} = c_k^* \left( \sum_{i=1, i\neq k}^{M} h_i h_i^* + \sigma^2 I \right)^{-1} c_k \tag{9}$$

On the other hand, we have

$$\begin{aligned} M_k &= c_k^* \left( h_k h_k^* + \sum_{i=1, i\neq u}^{M} h_i h_i^* + \sigma^2 I \right)^{-1} c_k \\ &= c_k^* \left( \sum_{i=1, i\neq k}^{M} h_i h_i^* + \sigma^2 I \right)^{-1} \left( I - \frac{h_k h_k^* \left( \sum_{i=1, i\neq k}^{M} h_i h_i^* + \sigma^2 I \right)^{-1}}{1 + h_k^* \left( \sum_{i=1, i\neq k}^{M} h_i h_i^* + \sigma^2 I \right)^{-1} h_k} \right) c_k \\ &= c_k^* \left( \sum_{i=1, i\neq k}^{M} h_i h_i^* + \sigma^2 I \right)^{-1} c_k - \frac{c_k^* \left( \sum_{i=1, i\neq k}^{M} h_i h_i^* + \sigma^2 I \right)^{-1} h_k h_k^* \left( \sum_{i=1, i\neq k}^{M} h_i h_i^* + \sigma^2 I \right)^{-1} c_k}{1 + h_k^* \left( \sum_{i=1, i\neq k}^{M} h_i h_i^* + \sigma^2 I \right)^{-1} h_k} \\ &= x_{-k} - \frac{x_{-k}^2 \left\| h_k \right\|^2}{1 + x_{-k} \left\| h_k \right\|^2} \\ &= \frac{x_{-k}}{1 + x_{-k} \left\| h_k \right\|^2} \\ &= \frac{x_{-k}}{1 + \gamma_k} \end{aligned} \tag{10}$$

We can see the metric $M_k$ is proportional to $\frac{1}{1+\gamma_k}$ with a varying coefficient $x_{-k}$ and thus the sequence possessing lower metric value procures in general data stream of higher post-SINR within the variable set if we acknowledge the following 2 assumptions. First is $\gamma_k$ dominates in terms of metric value variation and second is the variable set is restricted into the sequences actually selected by the UEs in lieu of the full selection pool.

In order to accommodate the variation of $x_{-k}$ as well as the fact that the variable set is the full selection pool, we constitute a candidate sequence pool of appropriate size $N_{cs}$ comprising the sequences whose metric values are the lowest ones to increase the probability of capturing the spread sequences procuring the data streams with highest post-SINR. This probability is dubbed as hit rate and will be evaluated later.

In autonomous grant-free high-overloading multiple access scenario where sequence collision is hard to avoid, for overloading as high as 400%, for example, 32 UEs multiplexed on 8 consecutive symbols occupied by the length-8 spread sequence, the hit rate stays over 90% per SIC round. The selection pool possesses 64 sequences and the candidate sequence pool comprises 8, thanks to which 7/8 equalization computational complexity is reduced  The blind sequence detection procedure is robust in the sense that even if the data stream with highest post-SINR is not picked out, an alternative sequence procuring  post-SINR superior to decoding threshold is likely to be captured within the candidate sequence pool. After the data stream is successfully decode, it could be reconstructed from the information bits from which the sequence information could be retrieved. This can be demonstrated by the two latter subfigures in Figure 3, where success rate is defined as the probability that at least one candidate sequence within the pool procures data stream succeeding the decoding procedure. Figure 3 illustrates the drop number distribution possessing diverse number of SIC rounds, and the total drop number is set as 1000. Similar evaluation results can be obtained for length 4 MUSA sequence.
Apart from reduced complexity, the blind sequence detection comes with following advantage. The procedure helps to maintain the success rate in case of deep fading where hard detection SINR doesn't comply with decoding SINR.

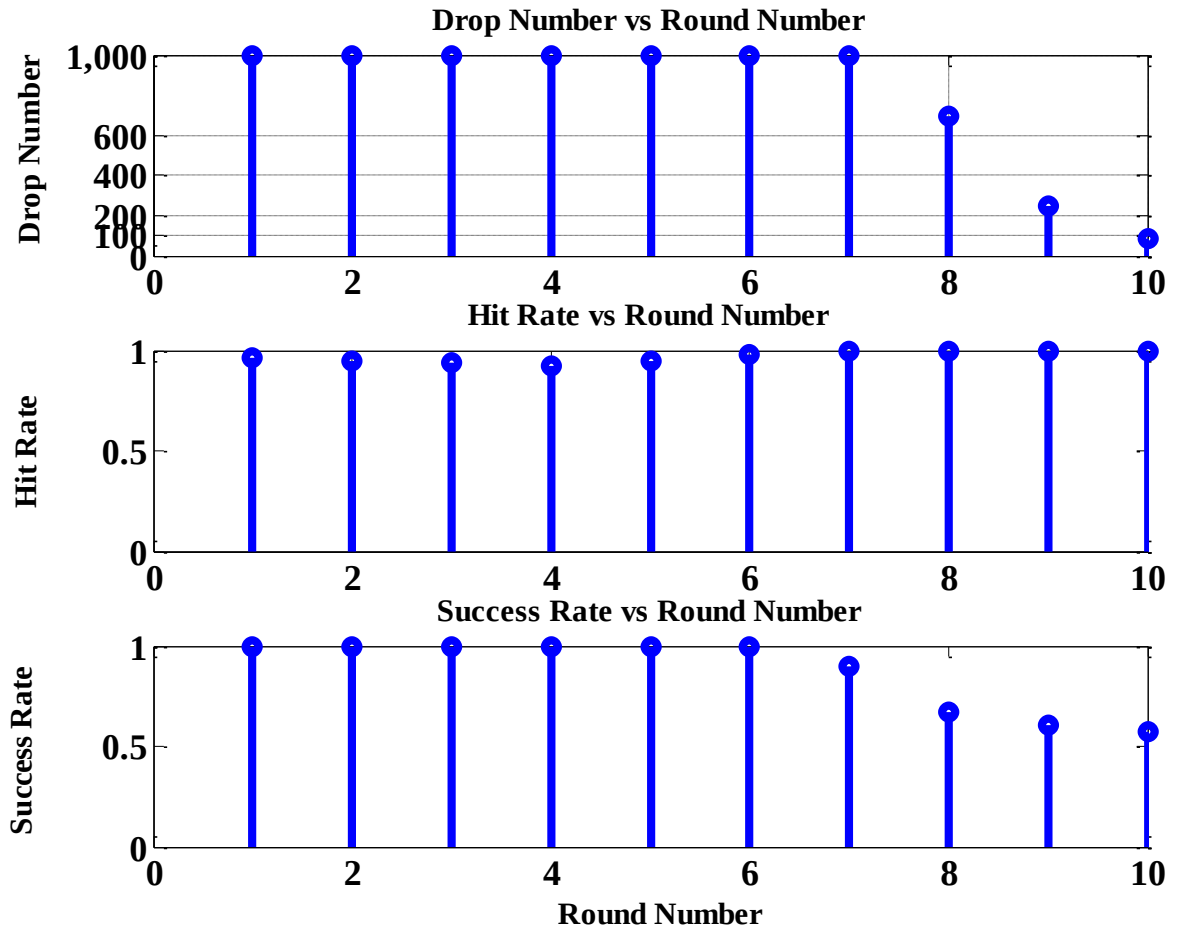

**Figure 3 Blind MMSE metric Evaluation**

### B. *Blind MMSE estimation*

Upon the initialization of each round of the SIC procedure, the reconstructed data streams from the previous rounds are subtracted from the  aggregate received signal leading to remaining spread unit-wise signal of higher SINR denoted as $y$

Assume there still exists $M$ undecoded UEs apart from UE $u$ for the SIC round considered, then the spread unit-wise MMSE estimation formula with equivalent channel information $h_u$ for UE $u$ is as follows,

$$\hat{x}_u = h_u^* \left( h_u h_u^* + \sum_{i=1,i\neq u}^{M} h_i h_i^* + \sigma^2 I \right)^{-1} y \qquad (11)$$

,where $h_u^*$ is the conjugate symmetry of $h_u$.

Data-only receiver has no access to the equivalent channel of any UE, making it hard if not impossible to directly apply the above equation to estimate the $\hat{x}_u$. On exploiting the following mathematical transformation, which is a direct result of (2) , we could replace the inverse matrix in the formula by the auto-correlation matrix of $y$.

$$yy^* = h_u x_u x_u^* h_u^* + h_u x_u n_u^* + n_u (h_u x_u)^* + n_u n_u^* \qquad (12)$$

On averaging both sides of the equation, we have

$$\begin{aligned} R_y &\overset{\Delta}{=} E(yy^*) \\ &= h_u h_u^* + h_u E(x_u n_u^*) + E(n_u x_u^*) h_u^* + E(n_u n_u^*) \\ &= h_u h_u^* + \sum_{i=1,i\neq u}^{M} h_i h_i^* + \sigma^2 I \end{aligned} \qquad (13)$$

It's worth noting that equation(13) is valid only if the channel stays invariant across all the spread units considered for averaging, which is a more reasonable assumption in the short spread sequence case.  On admitting the assumption that the sample average over all the spread units, denoted as $n_s$, approaches the statistical average and  the definition of equivalent channel in (6), we could replace the original MMSE estimation formula with the following one,

$$\hat{s}_u = g_u^* c_u^* R_y^{-1} y \qquad (14)$$

, where $R_y = \sum_{i=1}^{n_s} y_i y_i^* / n_s$

Considering the fact that $g_u^*$ is inaccessible to data-only receiver, we could obtain a pool of un-equalized blind MMSE estimated data streams $\hat{x}_{u_b}$ as follows by applying each candidate $c_u$ in the candidate sequence pool C from the blind sequence detection procedure.

$$\hat{x}_{u_b} = \left[ \frac{g_u \hat{s}_{u,1}}{\|g_u\|^2}, \frac{g_u \hat{s}_{u,2}}{\|g_u\|^2}, \dots, \frac{g_u \hat{s}_{u,n_s}}{\|g_u\|^2} \right] = c_u^* R_y^{-1} Y \qquad (15)$$

,where $Y = \left[ y_1, y_2, \dots, y_{n_s} \right]$ represents the received spread data stream.

The $\hat{x}_{u_b}$ constitutes two scatter point clusters originating from the BPSK modulated transmit symbols on the constellation map which will go through the blind equalization procedure followed by decoding operation elaborated in the later corresponding sections. Each element in $\hat{x}_{u_b}$ is noted as $\hat{s}_{u_b}$ in the following.

### *C. Blind Equalization*

The un-equalized data stream corresponds to 2 clusters of scatter points on the constellation map as illustrated in the first subplot in Figure 4 below. The original BPSK transmit symbols are illustrated in the second subplot on Figure 4. Taking full advantage of the geometric feature of the scatter point clusters, blind equalization component adopts a partition-summation technique to estimate channel and then conduct blind equalization whose principle and methodology will be elaborated in the following paragraphs. This component coupled with the component introduced in the previous section shall constitute the MMSE demodulation counterpart in the data-only receiver as that in conventional MMSE receiver.

Each scatter point on the constellation map corresponds to the spread unit-wise blind MMSE estimation result as follows,

$$c_u^* R_{yy}^{-1} y = \sum_{u \in U_c} \frac{g_u}{\|g_u\|^2} \hat{s}_u + \sum_{v \in V_I} \rho_v s_v + n \qquad (16)$$

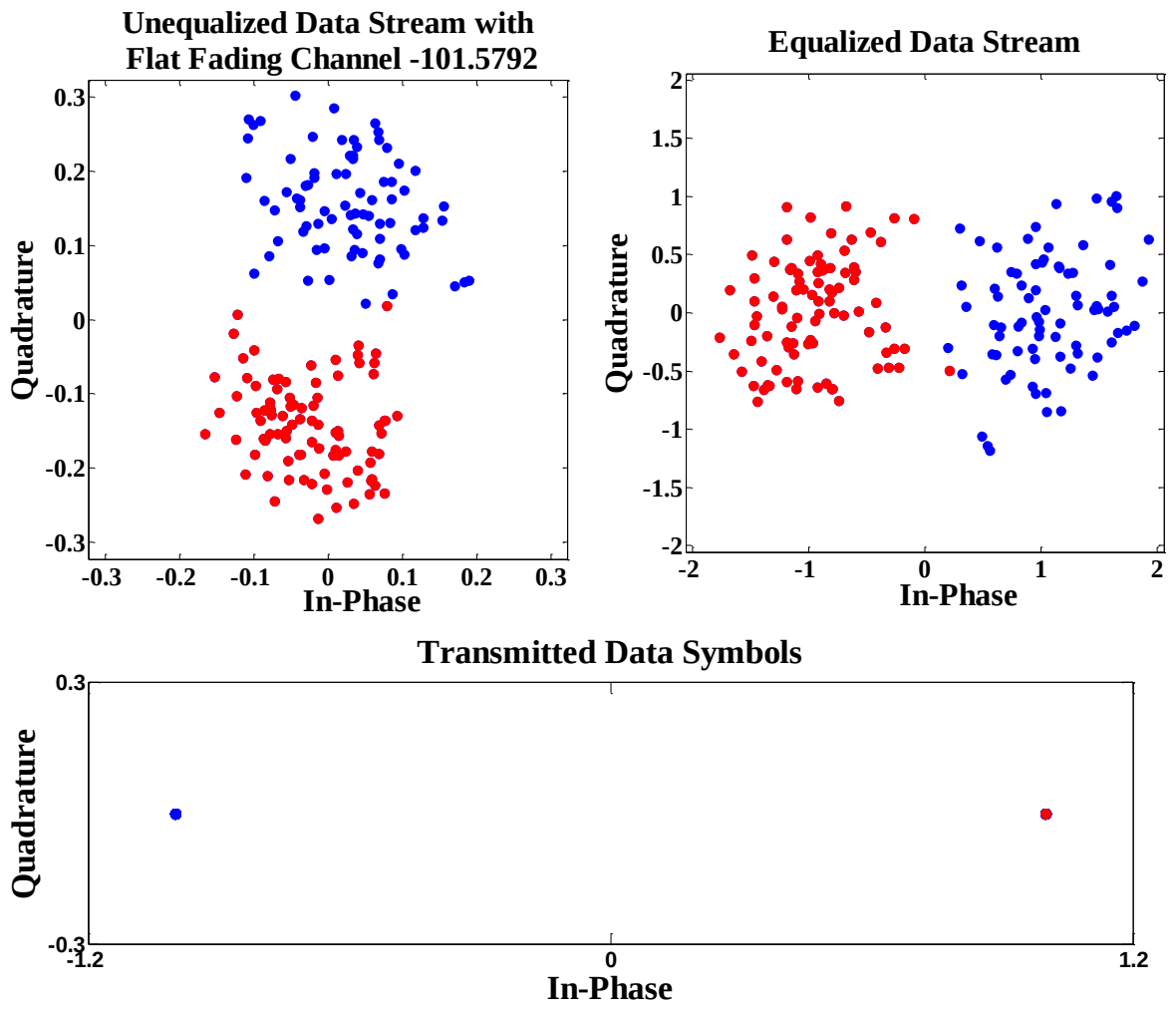

**Figure 4 Blind Equalization Illustration**

,where $\rho_v = c_u^* R_{yy}^{-1} h_v$ .and $n = c_u^* R_{yy}^{-1} n_{awgn}$ $U_c$ represents the set of collided UEs whose spread sequences are the same, and $V_I$ denotes the set consisting of UEs whose spread sequences differ from $c_u$. When it comes to the collision-free case, i.e., the number of UEs selecting the same spread sequence $c_u$ is 1, $U_c$ is reduced to a singleton. For simplicity of expression, we will first illustrate the principle and methodology regarding the blind equalization procedure by taking the collision-free case as an example and later extend our analysis to the collision case.

Denote $y_d = c_u^* R_{yy}^{-1} y$ and $\tilde{g}_u = \frac{g_u}{\|g_u\|^2}$, then we shall obtain the estimation of the physical channel $g_u$ as follows from the data streams $\hat{x}_{u_b}$,

$$g_u = \frac{\rho_u}{c_u^* R_y^{-1} c_u} \qquad (17)$$

, where $\rho_u \overset{\Delta}{=} E(s_u y_d)$. The proof of the formula could be found in appendix B. The key of blind equalization procedure lies in obtaining the estimation of $g_u$ and compensate its conjugate to the unequalized data stream $\hat{x}_{u_b}$ from the previous section. According to the above relation, $\rho_u$ can be approximated by the sample average of the product of spread unit-wise blind MMSE estimated symbol and the original BPSK transmit symbol across all the spread units. This motivates us to enforce the blind equalization procedure via the following 2 steps. First, divide the scatter points on the constellation map into 2 halves corresponding respectively to opposite transmit BPSK

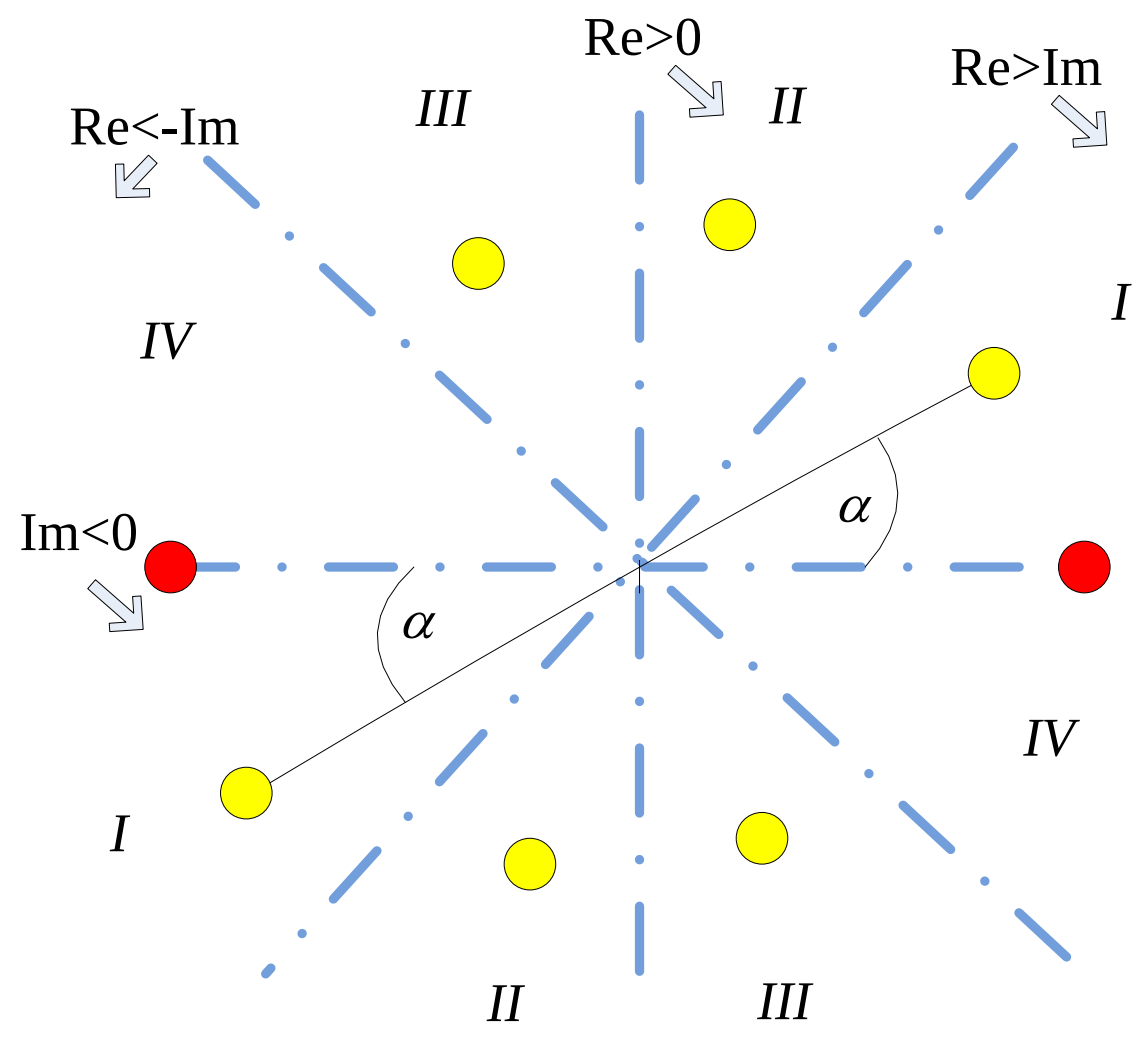

**Figure 5 Partition of the Complex Plane**

symbols{+1, -1}. Second sum up and average the scatter points of the 2 sections multiplied by their transmit symbol. Both steps will be elaborated in the following.

As shown in the Figure 5 above, the un-equalized data stream $\hat{x}_{u_b}$ corrupted with the interference from the other UEs as well as the AWGN is scattered on the constellation map above and highly likely to reside in either or multiple of the 4 regions depending on the post SINR. With 2 pairs of boundary, respectively the x axis and y axis in the coordinate system and the 45-degree and 135-degree boundary, the plan is divided into 8 pieces each 2 of which make up a region accommodating respectively the scatter point clusters originating from the BPSK transmit symbols +1 and -1. We see from Figure 5 that for the scatter point clusters located in regions II&III, the x axis functions as the best boundary in the sense that it is perpendicular to the line connecting the centers of both clusters. Likewise, the scatter point clusters in regions I & II correspond to the 135 degree boundary. And the scatter points in regions III&IV correspond to the 45 degree boundary as illustrated in the constellation map. The scatter points in region I&IV correspond to the y axis boundary. The 4 boundaries constitutes an isotropic division of the 2-dimensional plan and thus covers to a great extent diverse shapes and distributions of scatter point clusters.

Each of the 4 boundaries divide the constellation map into 2 regions, $R_1$ and $R_{-1}$. The scatter point cluster in $R_1$ is regarded as originated from transmit symbol 1 and thus is assigned weight 1 while the scatter point cluster in $R_{-1}$ is -1. It's worth noting that the assigned weight could be the inverse of the actual transmit symbol, as cluster in $R_1$ could also originate from transmit symbol -1. This uncertainty causes a phase ambiguity problem which will be treated later.

Thanks to the previous division and weight assignment, the sample average of the product of spread unit-wise blind MMSE detected symbol and the original BPSK transmit symbol across all the spread units is obtained as follows,

$$\hat{x}_{u_c} = \frac{\left(\sum_{R_1} \hat{s}_{u_b} - \sum_{R_{-1}} \hat{s}_{u_b}\right)}{n_s} \quad (18)$$

Out of the 4 boundaries, the one bringing about $\hat{x}_{u_c}$ with the largest module is supposed to have divided the plans into halves best accommodating the clusters originating from transmit symbols or the opposites, and the phase of this $\hat{x}_{u_c}$ , denoted as $\theta_{u_c}$ is compensated to the blind MMSE estimated data stream $\hat{x}_{u_b}$ ,

$$\hat{x}_{u_d} = \hat{x}_{u_b} \cdot e^{-j\theta_{u_c}} \quad (19)$$

$\hat{x}_{u_d}$ is ready for decoding.

To cope with the phase ambiguity problem, both $\hat{x}_{u_d}$ and $-\hat{x}_{u_d}$ are delivered to the decoder and once either data stream $\hat{x}_{u_d}$ passes the CRC check, the corresponding decoded data will be reconstructed and go through the SIC procedures. As a matter of fact, the blind equalized data stream in Figure 4 suffers from the phase ambiguity problem and $-\hat{x}_{u_d}$ instead of $\hat{x}_{u_d}$ will succeed the CRC check.

When it comes to the collision case, the blind equalization technique could still retrieve the channel of the UE with the highest post-SINR among the UEs spread by collided sequences., thanks to which the decoding SINR of the UE is further improved. Though the UEs are superposed in the spread sequence domain, the low correlation of the their transmit data helps to ensure the accuracy of the data-pilot based joint channel estimation in the following section.

The transmit signal from 2 UEs sharing the same spread sequence yet possessing different physical channel is illustrated in the constellation map in the first subfigure in Figure 6, its blind MMSE estimation result omitting the interferences is illustrated in the second subfigure We see the blind equalization technique could obtain the channel phase of the UE with stronger channel gain, i.e., higher post-SINR, by calculating the phase of the center of the scatter points connected by the red dotted lines. The boundary dividing the plan is therefore the y axis.

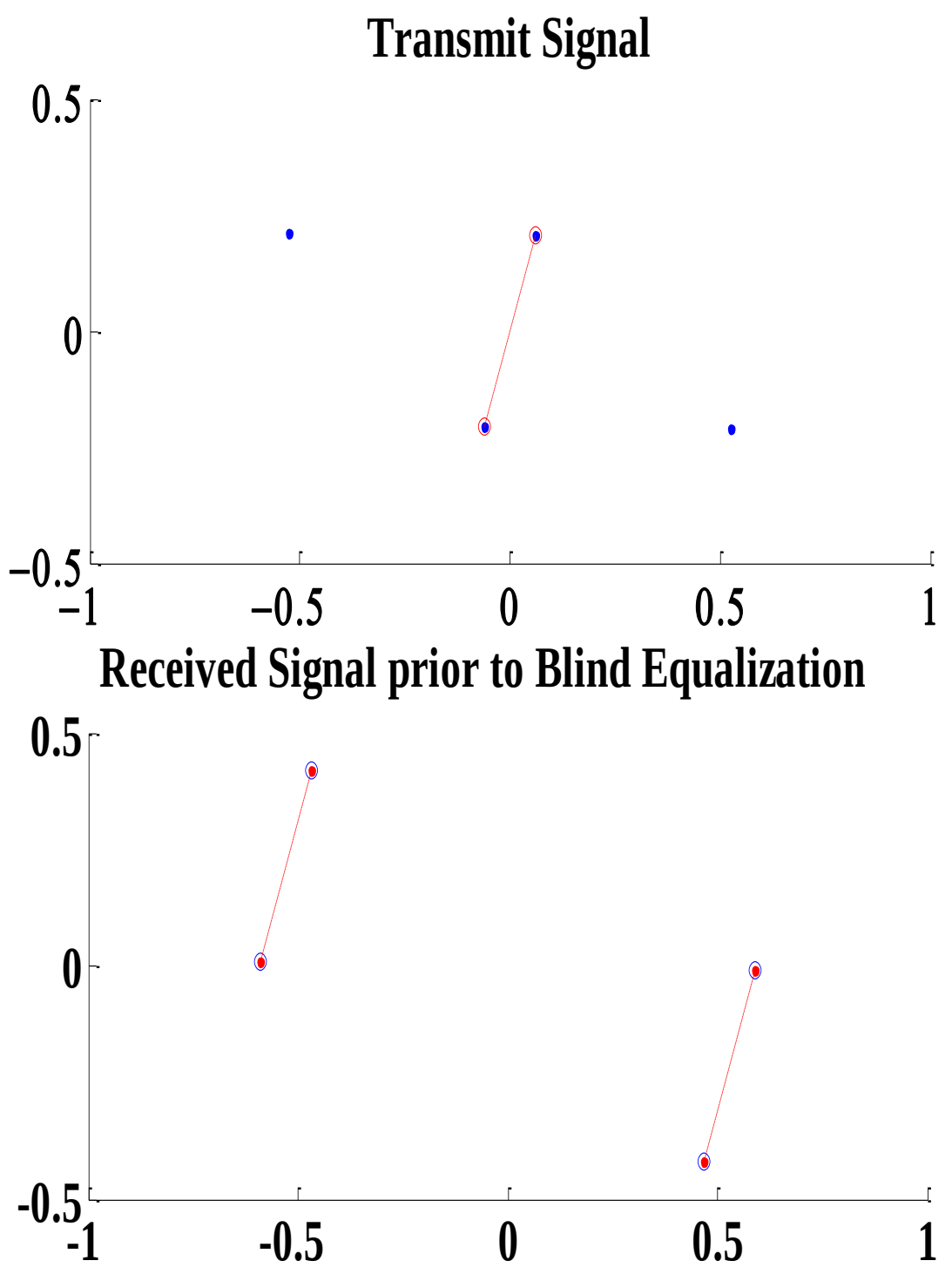

**Figure 6 Two UE Collision Demonstration**

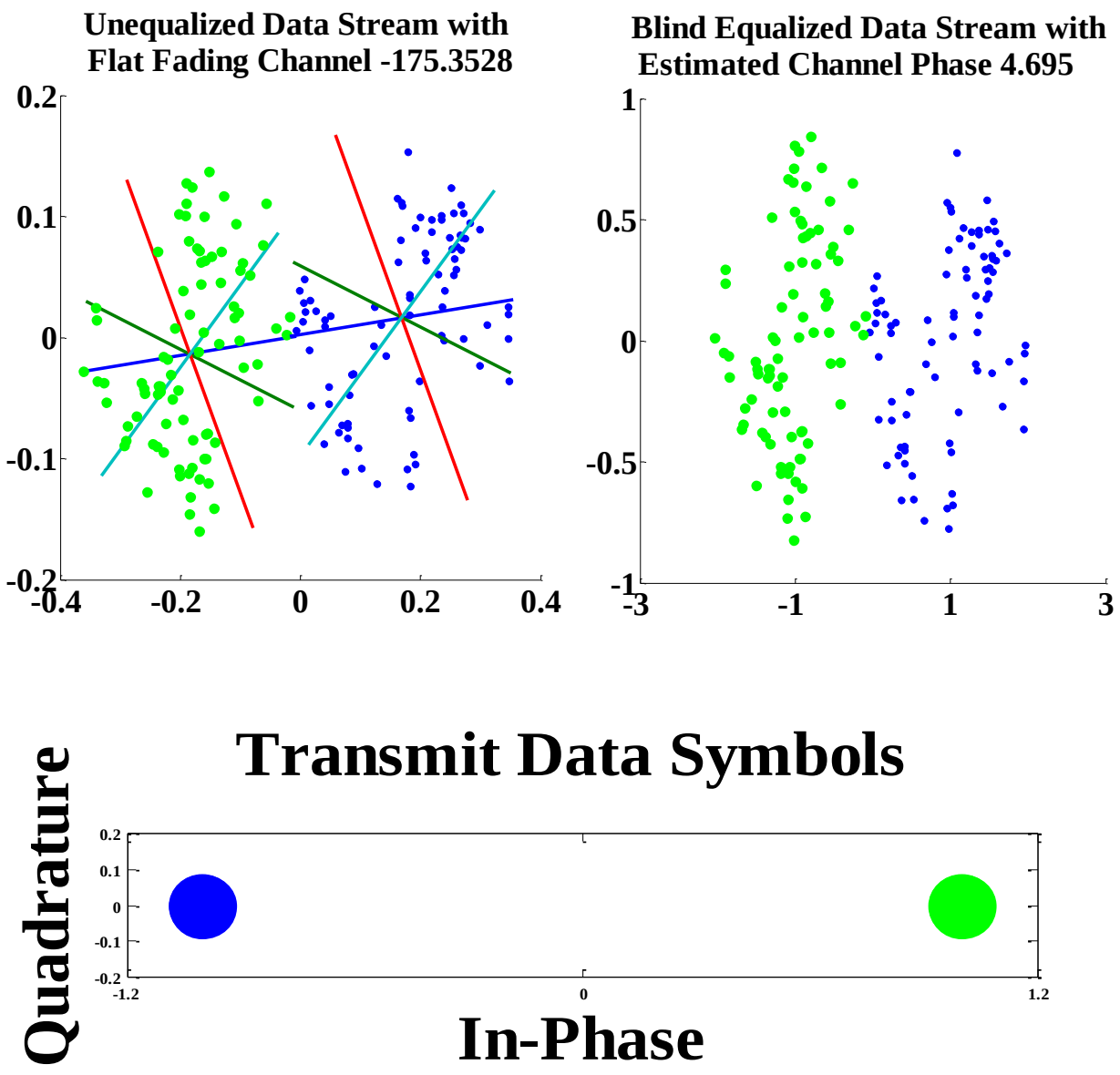

**Figure 7 Blind Equalization in Collision Case**

Figure 7 illustrates the case where the collided UE numbers are as high as 4 and their transmit SNR ranges from 4~20. The UE with the highest SINR has the channel direction shown by the blue line in the figure. The equalized data stream shown in the first two subfigures suffers from a phase ambiguity and its inverse succeeds the CRC check.

### *D. Data pilot channel estimation*

Subject to the constraint of computational complexity, part of the equalized data streams with highest hard detection SINR are passed on to the decoding process. For a given SIC round, once the UEs are decoded, the decoded UE data is employed to further update the channel estimations of the UEs decoded in the previous rounds In this way, the UE channels are iteratively refined, based on which the data streams from previous rounds are reconstructed again minimizing the residual error in the data stream for this SIC round. Assuming the cumulative decoded UE number is $Q$, we have the following equation,

$$y=\left[s_1,s_2,\ldots,s_Q\right]\left[h_1,h_2,\ldots,h_Q\right]^t+n_Q \qquad (20)$$

where $s_i$ is the decoded data of UE $i$ within the range of the coherence bandwidth and time and $h_i$ represents the static scalar channel. $n_Q$ is the remaining interference and noise. In the simulation, we set the coherence range as the whole PRB as flat fading channel is assumed.

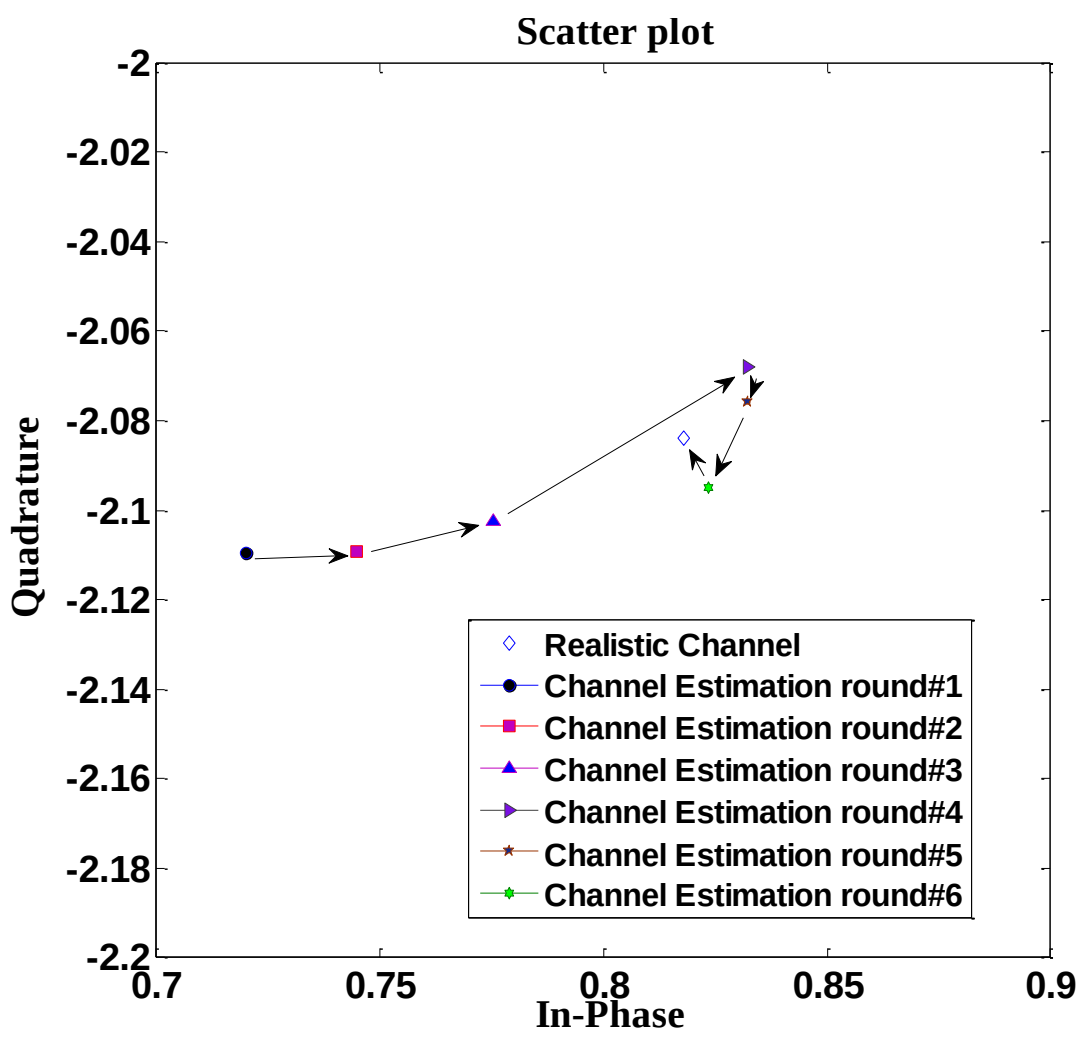

**Figure 8 Data-pilot Aided Channel Estimation**

On applying the least-square(LS) metric to all the decoded data till the end of this round, the UE wise estimation of the channel is calculated as follows,

$$\left[h_1,h_2,\ldots,h_Q\right]^t=\begin{pmatrix} s_1^*s_1 & \cdots & s_1^*s_Q \\ \vdots & \ddots & \vdots \\ s_Q^*s_1 & \cdots & s_Q^*s_Q \end{pmatrix}^{-1}\left[s_1^*y,s_2^*y,\ldots,s_Q^*y\right]^t \qquad (23)$$

As shown in Figure 8, the channel estimation accuracy improves as the channel updates with more decoded data employed in estimation.

## IV. PERFORMANCE

To evaluate the performance of the data-only receiver in combination with the code set in Appendix A in high-overloading AGF scenario. The spread sequence of each UE is assumed to be randomly selected. Link level simulations adopting the following parameters are carried out. Single transmit antenna and receive antenna is assumed. Legacy LTE resource configuration parameters are employed and the MCS is set to be turbo coding with 1/2 code rate and BPSK modulation. UEs are assumed to possess transmit SNR within a range of 8dB around the central SNR, as a result of the open-loop power control in AGF scenario. Flat channel fading is assumed. It should be noted that with the variation of parameters such as the size of blind detected sequence pool or the blind equalized data streams delivered to decoding, performance fluctuations could be expected, in particular in scenarios with To/Fo. In the following simulations, the numbers are set to be 10 and 5 respectively, based on which the BLER variations with respect to SNR are evaluated.

In addition to the high-overloading and the perceivable gap to the ideal case where ideal channel estimation is assumed.

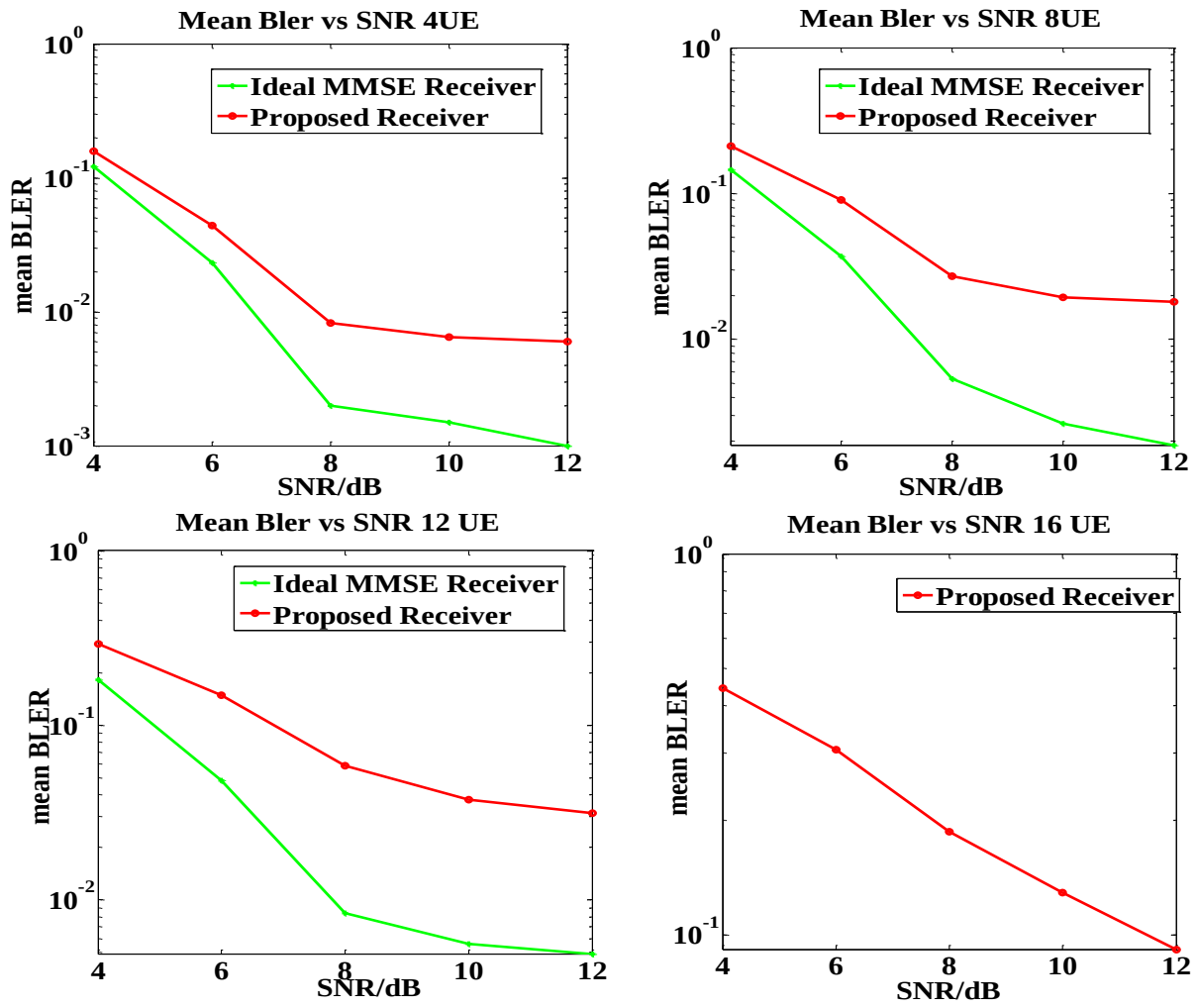

**Figure 9 Performance Evaluation for 4-16 UEs**

Thanks to the reduced RS overhead, per UE SE is as high as 0.125bits/s/Hz, which is more spectrally efficient than schemes relying on DMRS to perform channel estimation. The spectral efficiency further increases as the system loading increases. From the above Figure 8, 16 UEs whose SNR uniformly distributed in the range of 4~20 dB could be supported.

# V. Conclusions

In this paper, design challenges as well as potentials are discussed for the high-overloading AGF transmission scenario. Short spread sequence based one-shot transmission coupled with data-only receiver framework is promising in terms of transforming the potentials into realistic system gains. To this end, a highly efficient blind multiple user detection(MUD) for AGF multiple access without any reference signal is described and evaluated given the MUSA sequence listed in Appendix A. The blind MUD targets the spreading-based high-overloading grant-free scenario. Of course, for the less challenging data-only grant-free cases such as light loading system, or non-spreading based, and even the grant-based data transmission without reference signals, some 'blind' ideas in this paper can also be applied, underpinning an implementation-friendly and performance-superior blind MUD receiver. Further study is necessary in multiple-antenna scenarios where efforts should be paid to take full advantage of the antennas to deliver additional system gains. Some of ideas in blind equalization techniques could also be applied to scenarios with Tx/Rx impairments.

# VI. Appendix

## A. *MUSA Code Set*

SF=4, pool size =64(before normalization)

| No. | C1 | C2 | C3 | C4 | No. | C1 | C2 | C3 | C4 |
|---|---|---|---|---|---|---|---|---|---|
| 1 | 1 | 1 | -1 | -1 | 33 | 1 | -j | j | -j |
| 2 | 1 | j | -j | 1 | 34 | 1 | -1 | -j | -1 |
| 3 | 1 | j | j | -1 | 35 | 1 | 1 | j | -1 |
| 4 | 1 | 1 | 1 | 1 | 36 | 1 | j | 1 | 1 |
| 5 | 1 | -1 | -j | -j | 37 | 1 | j | 1 | -1 |
| 6 | 1 | -j | 1 | j | 38 | 1 | -j | j | j |
| 7 | 1 | 1 | -j | j | 39 | 1 | j | j | j |
| 8 | 1 | -1 | 1 | -1 | 40 | 1 | -1 | -1 | j |
| 9 | 1 | -j | -1 | -j | 41 | 1 | j | -1 | -1 |
| 10 | 1 | -j | j | 1 | 42 | 1 | 1 | j | 1 |
| 11 | 1 | 1 | j | -j | 43 | 1 | 1 | 1 | -j |
| 12 | 1 | j | 1 | -j | 44 | 1 | 1 | -1 | -j |
| 13 | 1 | -1 | j | j | 45 | 1 | j | -j | -j |
| 14 | 1 | j | -1 | j | 46 | 1 | -1 | -1 | -j |
| 15 | 1 | -j | -j | -1 | 47 | 1 | -j | -1 | 1 |
| 16 | 1 | -1 | -1 | 1 | 48 | 1 | -j | -j | j |
| 17 | 1 | -1 | j | -j | 49 | 1 | -1 | 1 | -j |
| 18 | 1 | 1 | -j | -j | 50 | 1 | -1 | j | 1 |
| 19 | 1 | -j | 1 | -j | 51 | 1 | j | j | -j |
| 20 | 1 | -1 | 1 | 1 | 52 | 1 | 1 | -j | 1 |
| 21 | 1 | 1 | 1 | -1 | 53 | 1 | j | -1 | 1 |
| 22 | 1 | -1 | -j | j | 54 | 1 | 1 | -j | -1 |
| 23 | 1 | j | -1 | -j | 55 | 1 | -1 | -j | 1 |
| 24 | 1 | j | 1 | j | 56 | 1 | -1 | j | -1 |
| 25 | 1 | j | -j | -1 | 57 | 1 | -j | 1 | -1 |
| 26 | 1 | -j | j | -1 | 58 | 1 | -1 | 1 | j |
| 27 | 1 | -j | -1 | j | 59 | 1 | 1 | -1 | j |
| 28 | 1 | -1 | -1 | -1 | 60 | 1 | -j | 1 | 1 |
| 29 | 1 | -j | -j | 1 | 61 | 1 | 1 | 1 | j |
| 30 | 1 | 1 | -1 | 1 | 62 | 1 | j | -j | j |
| 31 | 1 | j | j | 1 | 63 | 1 | -j | -j | -j |
| 32 | 1 | 1 | j | j | 64 | 1 | -j | -1 | -1 |

## B. *Blind MMSE Estimation and Equalization*

We know from (16) that to obtain the MMSE estimation $\hat{s}_u$ of the signal , a multiplication of $g_u^*$ to the left of the equation suffices. To estimate the $g_u^*$ from the un-equalized data stream, we resort to the data-only transceiver system model (2) and the fundamentals of MMSE estimation.

Conventional MMSE estimation of UE $u$ with equivalent channel $h_u$ can be decomposed into 3 steps, first whiten the colored noise by its co-variance matrix $K_{n_u}^{-1/2}$, then scale the result with $\frac{1}{1+h_u^* K_{n_u}^{-1} h_u}$, and then match filter with

$K_{n_u}^{-1/2} h_u$ .[25]

We can rewrite the spread unit-wise transceiver system model prior to matched filter $v = K_{n_u}^{-1/2} h_u$ as follows,

$$\tilde{y} = \frac{K_{n_u}^{-1/2} y}{1+h_u^* K_{n_u}^{-1} h_u} = \frac{K_{n_u}^{-1/2} h_u s_u + K_{n_u}^{-1/2} n_u}{1+h_u^* K_{n_u}^{-1} h_u}$$

Instead of the matched filter, blind MMSE estimation applies the following filter to the above whitened received signal. This procedure is followed by the multiplication of the original transmit signal $s_u$ in the blind equalization component,

$$v = K_{n_u}^{-1/2} c_u$$

$c_u$ is the spread sequence selected by UE u.

The resultant signal of the blind equalization is given as follows

$$s_u v^* \tilde{y} = \frac{c_u^* K_{n_u}^{-1} h_u}{1+h_u^* K_{n_u}^{-1} h_u} + \frac{c_u^* K_{n_u}^{-1} n_u s_u}{1+h_u^* K_{n_u}^{-1} h_u}$$

Denote $\frac{c_u^* K_{n_u}^{-1} n_u s_u}{1+h_u^* K_{n_u}^{-1} h_u}$ as $n_b$

As $n_b$ is the linear combination of independently distributed zero-mean Gaussian random variables, and $n_b$ is independent from $s_u$, we have

$$E(n_b s_u) = E(n_b) E(s_u) = 0$$

Consequently, we have

$$E\left(s_u v^* \tilde{y}\right) = \frac{c_u^* K_{n_u}^{-1} h_u}{1+h_u^* K_{n_u}^{-1} h_u}$$

Denote it as $\rho_u$ ,

We have $g_u = \frac{\rho_u}{c_u^* R_y^{-1} c_u}$ and the deductions are introduced as follows

$$\rho_u = \frac{c_u^* \left( \sum_{i=1, i\neq u}^{M} h_i h_i^* + \sigma^2 I \right)^{-1} h_u}{1 + h_u^* \left( \sum_{i=1, i\neq u}^{M} h_i h_i^* + \sigma^2 I \right)^{-1} h_u}$$

$$= c_u^* \left( \sum_{i=1, i\neq u}^{M} h_i h_i^* + \sigma^2 I \right)^{-1} h_u - \frac{c_u^* \left( \sum_{i=1, i\neq u}^{M} h_i h_i^* + \sigma^2 I \right)^{-1} h_u h_u^* \left( \sum_{i=1, i\neq u}^{M} h_i h_i^* + \sigma^2 I \right)^{-1} h_u}{1 + h_u^* \left( \sum_{i=1, i\neq u}^{M} h_i h_i^* + \sigma^2 I \right)^{-1} h_u}$$

$$= c_u^* \left( \sum_{i=1, i\neq u}^{M} h_i h_i^* + \sigma^2 I \right)^{-1} \left( I - \frac{h_u h_u^* \left( \sum_{i=1, i\neq u}^{M} h_i h_i^* + \sigma^2 I \right)^{-1}}{1 + h_u^* \left( \sum_{i=1, i\neq u}^{M} h_i h_i^* + \sigma^2 I \right)^{-1} h_u} \right) h_u$$

$$= c_u^* \left( \sum_{i=1, i\neq u}^{M} h_i h_i^* + \sigma^2 I \right)^{-1} h_u - \frac{c_u^* \left( \sum_{i=1, i\neq u}^{M} h_i h_i^* + \sigma^2 I \right)^{-1} h_u h_u^* \left( \sum_{i=1, i\neq u}^{M} h_i h_i^* + \sigma^2 I \right)^{-1} h_u}{1 + h_u^* \left( \sum_{i=1, i\neq u}^{M} h_i h_i^* + \sigma^2 I \right)^{-1} h_u}$$

$$= \frac{c_u^* \left( \sum_{i=1, i\neq u}^{M} h_i h_i^* + \sigma^2 I \right)^{-1} h_u}{1 + h_u^* \left( \sum_{i=1, i\neq u}^{M} h_i h_i^* + \sigma^2 I \right)^{-1} h_u}$$

$$= c_u^* R_y^{-1} h_u$$

Here we encounter the blind MMSE metric again in the denominator $c_u^* R_y^{-1} c_u$ , whose module is inversely correlated with that of $g_u$ ,providing another implicit explanation on the metric's relation to the post-SINR of the MMSE demodulated data stream .